\documentclass{article}
\usepackage[a4paper,top=1cm,bottom=1cm,left=1.2cm,right=1.2cm]{geometry}

\usepackage{graphicx}
\usepackage{dcolumn}
\usepackage{bm}
\usepackage{amsmath}
\usepackage{lineno}
\usepackage{mathtools}
\usepackage[section]{placeins}
\usepackage{textcomp}
\usepackage{authblk}
\usepackage{xcolor}
\usepackage{braket}
\usepackage{amssymb}

\begin{document}

\title{Efficient Quantum Space-Division Multiplexing Using Time-Bin and Phase Encoding in Few-Mode Fibers}

\author[1*]{Mario Zitelli}

\affil[1]{Department of Information Engineering, Electronics, and Telecommunications, Sapienza University of Rome, Via Eudossiana 18, 00184 Rome, Italy}
\affil[*]{mario.zitelli@uniroma1.it}

\maketitle

 \begin{abstract}
 Space-division multiplexing using multimode optical fibers has been applied to quantum-level signals with time-bin and phase encoding, achieving Mqubits per second over 8 km of few-mode fiber. The dead time of single-photon detectors, typically considered a bottleneck in quantum transmission, is effectively utilized here to eliminate modal cross-talk between quantum channels.
 \end{abstract}

\section{Introduction}

The transmission of cryptographic keys between two parties in such a way that no amount of computational power—classical or quantum—supported by any algorithm \cite{doi:10.1137/S0097539795293172} can retrieve any information about them, is the purpose of Quantum Key Distribution (QKD). In recent years, QKD has seen remarkable progress, enabling secure quantum communication over distances up to 600 km \cite{pittaluga2021600} with key rates exceeding 10 Mbps, or gigabit bandwidth over 60 km \cite{Yu2025}.
Traditionally, single-mode fibers are employed for long-distance quantum key transmission \cite{PhysRevLett.126.250502,10.1063/1.5027030}. However, both multicore fibers (MCFs) \cite{Dynes:16,bacco2019boosting,9102386,xavier2020quantum} and multimode fibers (MMFs) \cite{Namekata:05,Amitonova:20,Wang:20,Zhong:21} have been explored for QKD, as part of efforts to transmit classical and quantum signals simultaneously in different spatial modes of the same fiber, using the spatial-division multiplexing (SDM) technique \cite{Winzer:18}.

Quantum transmission capacity can be further enhanced through multilevel QKD, where each optical pulse carries multiple qubits. This is achieved by encoding qubit information into various degrees of freedom such as optical phase, pulse position, polarization, or spatial modes.

In \cite{Inoue_09}, the relative phase between consecutive single-photon pulses was used to implement the BB84 protocol \cite{Nielsen_Chuang_2010}. The transmitter, Alice, prepares four qubits by modulating the differential phase of the pulses using one of two bases: $\phi_{A1} = \{0, \pi\}$ or $\phi_{A2} = \{\pi/2, 3\pi/2\}$, chosen at random, according to a bit sequence $a = \{a_i\}$ and a basis sequence $b = \{b_i\}$. The pulses are attenuated to below one photon per pulse. The receiver, Bob, applies a phase modulation $\phi_B = \{0, \pi/2\}$ to adjacent pulses according to his basis sequence $b' = \{b'_i\}$. After raw key transmission, Alice and Bob reveal their basis choices and detection times over a public channel, allowing them to sift the transmitted qubits and establish a shared secret key. In the work, pulses were detected at a gate rate of 1 MHz.

In \cite{Ecker_2019}, time-energy entanglement within the coherence time of a pump laser was employed to transmit two-photon states. The spontaneous parametric down-conversion (SPDC) process, using a coherent laser and a nonlinear crystal (ppKTP), generates photon pairs with time uncertainty $\Delta t$ due to the Heisenberg uncertainty principle. For lasers with large coherence times $t_p$, this results in entanglement in the time domain. Specifically, photon pairs transmitted at times $\ket{t}$ and $\ket{t+\tau}$ within the coherence window form a two-photon state $\ket{\psi} = \ket{t} \otimes \ket{t+\tau}$. Coincidences were detected at 1 kHz rate.

Reference \cite{Steinlechner_2017} proposed multilevel hyper-entanglement in polarization and time-energy degrees of freedom, combining linear polarization states $\ket{H}$, $\ket{V}$ with time-energy positions $\ket{t}$, $\ket{t+\tau}$ to form a four-dimensional state $\ket{\psi}=1/2(\ket{H}_A\ket{H}_B+\ket{V}_A\ket{V}_B) \otimes (\ket{t}_A\ket{t}_B+\ket{t+\tau}_A\ket{t+\tau}_B)$, where $A$ and $B$ are SPDC-generated photon pairs, with same or different wavelengths. These four-dimensional states were successfully transmitted over a 1.2-km free-space link, with 20 kcps two-photon detection events.

In \cite{PhysRevLett.122.020503}, multilevel orbital angular momentum (OAM) states were proposed, with photons prepared in superpositions of OAM modes and transmitted through free space. At the receiver, OAM information was decoded using a spatial light modulator (SLM). Qudits with $d = 32$ levels—equivalent to $\log_2 d = 5$ qubits—were successfully transmitted.

Reference \cite{doi:10.1126/sciadv.1701491} presented a time-bin and phase multilevel modulation scheme, where optical pulses containing one photon or less are transmitted with controlled time delays. Information is encoded in the delay of the pulse $\ket{t_k}$ into its time slot $\Delta t=d/R$, with $R$ being the pulse repetition rate. Time delays are detected by a single-photon sensor $D_t$. To verify channel integrity and detect eavesdropping, a phase state composed of a coherent pulse train at rate $R$ is alternated with the delayed pulse. The phase state is received using a series of time-delay interferometers and single-photon detectors. Phase fluctuations induced by an eavesdropper would be revealed in the phase state. The work demonstrated transmission of $d=4$ pulse trains with $R=2.5 GHz$ and detection rate $< 1-2$ Mcps.

Multilevel time-bin modulation was recently improved in \cite{vijayadharan2025sagnacbasedarbitrarytimebinstate}, where a distributed feedback (DFB) laser and two Sagnac interferometer-based intensity-phase modulation stages were used for pulse carving and picking. Four-dimensional time-bin states were transmitted at a frame rate of 40 MHZ, with frames composed by a train of phase-modulated pulses.

In this work, we revisit the time-bin and phase encoding approach described in \cite{doi:10.1126/sciadv.1701491} to demonstrate the feasibility of further multiplexing quantum information in an SDM system. The definition of time-bin in our work refers to the generation of frames containing single pulses, whose arrival time is measured, alternated with the generation of pulse trains with phase modulation, used for channel analysis.

We simultaneously transmitted three quantum channels through 8 km of multimode fiber, each carrying multilevel information with $d = 64$ (6 qubits per pulse). The resulting aggregated quantum transmission capacity reached 1.22 Mqubits/s on a single wavelength, illustrating how SDM, combined with multilevel encoding, can enhance quantum transmission respect to single-mode fiber QKD, and is compatible with other multiplexing techniques such as wavelength-division multiplexing (WDM) \cite{Zia:10799905}.

This was enabled by the use of multiplane light conversion (MPLC) \cite{Bade:18}, a recent technique for multiplexing and demultiplexing quantum information into spatial modes of an MMF. An MPLC-based spatial multiplexer \cite{fontaine2019} transforms $M$ Gaussian beams from single-mode input fibers into $M$ distinct propagation modes of an MMF—such as Laguerre-Gauss (LG), Hermite-Gauss (HG), or other modal bases used in free-space optics. MPLC has been shown to support up to 45 HG or LG modes in a standard 50 $\mu$m core graded-index MMF. The same MPLC setup also performs the inverse operation, enabling complete multiplexing and demultiplexing of multiple spatial channels with relatively low insertion losses, which is critical in quantum optical systems.

In our system, the dead time of single-photon detectors—typically required to dissipate the accumulated charge after each detection—was strategically utilized to mitigate modal cross-talk between signals. The phase states we employed are crucial for identifying potential eavesdroppers, through the detection of phase fluctuations between consecutive pulses. Alternatively, a BB84 protocol supported by phase states can be implemented—not for information transmission—but as a robust detection mechanism against eavesdropping.

\section{\label{sec:ExpResults}Experimental Results}

\begin{figure*}
\centering
\includegraphics[width=0.8\textwidth]{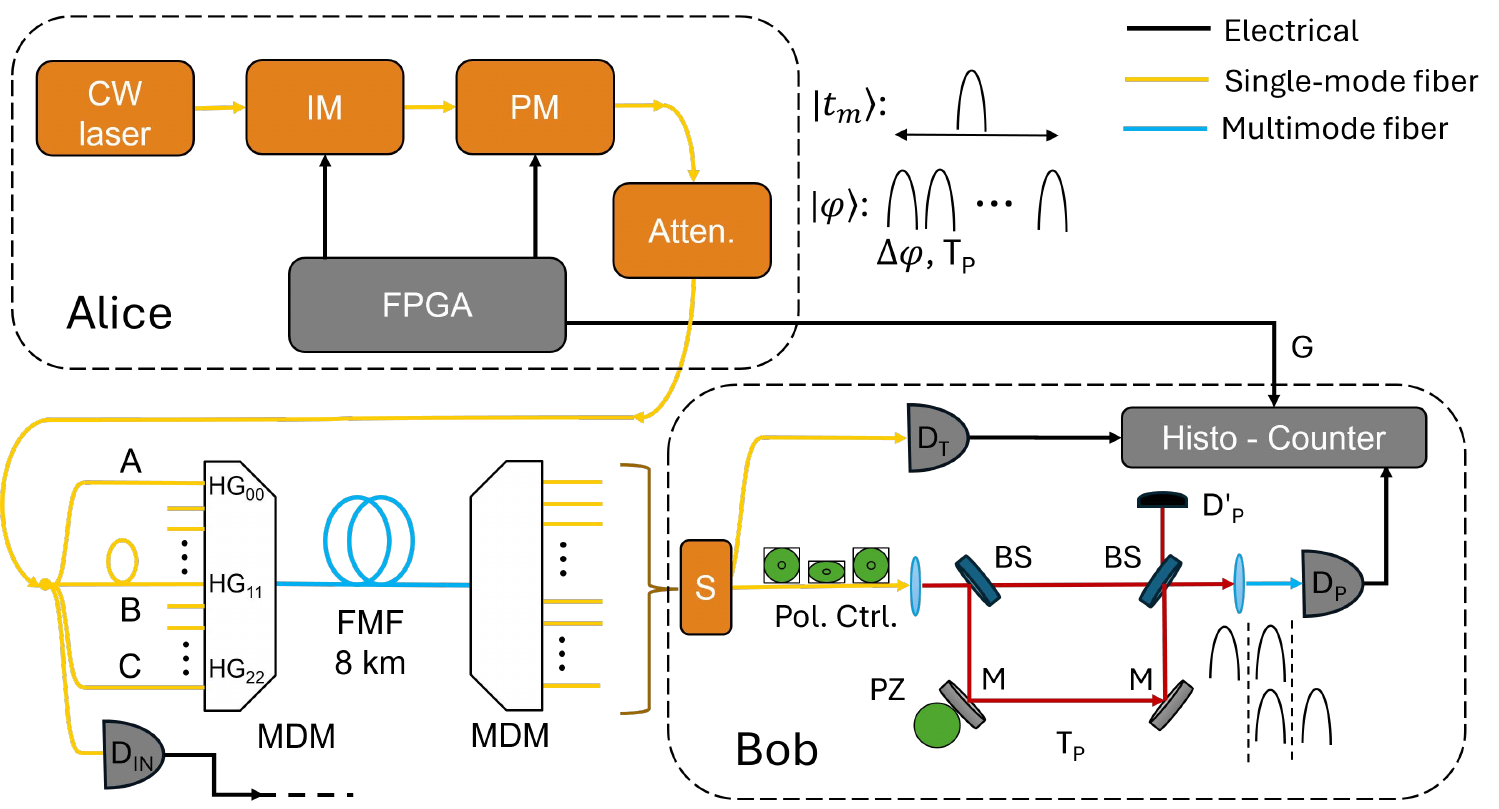}
\caption{The SDM transmision setup}
\label{fig:Setup}
\end{figure*}

The setup used for the transmission experiment is shown in Fig. \ref{fig:Setup}. It includes a multilevel transmitter (Alice), a transmission line consisting of 8 km of multimode fiber with modal multiplexers, and a multilevel receiver (Bob).

Alice consists of a continuous-wave (CW) laser, tunable in the C-band around $\lambda_0 = 1550$ nm with a 10 kHz linewidth (Thorlabs TLX1), followed by an optical intensity modulator (IM, Exail MX-LN-10) and a phase modulator (PM, Thorlabs LN-65S), both driven by a programmable FPGA with 8 GHz bandwidth (Zynq UltraScale + RFSoC ZCU670). The IM generates a frame composed of a sequence of $d = 64$ pulses with a pulse period of $T_p = 1.54$ ns. The 64 pulses span a frame duration $\Delta t = 100$ ns, followed by a blank interval of equal duration, for reasons that will be discussed shortly. Pulses are generated by driving the IM in a push-pull configuration and biasing it at the minimum of its transmission curve to maximize extinction ratio. Consequently, adjacent pulses experience a phase shift $\Delta\phi = \pi$. The PM adds a synchronous phase modulation, producing an output sequence of $d = 64$ pulses with an arbitrary phase difference $\phi_A$ between adjacent pulses. Signals are then attenuated to a level of $\mu$ photons per $d$-pulse frame, with $\mu$ measured at the input of the modal multiplexer. The resulting phase state is:

\begin{equation}
\ket{\phi} = \sum_{m=1}^{d} \frac{\ket{t_m} \exp(im\phi_A)}{\sqrt{d}}
\label{eq:Phase_state}
\end{equation}

Alternatively, a single pulse is generated with a delay $t_m$ within the frame. A total of 64 pulse positions are available in the 100 ns frame, corresponding to 6 qubits per frame. The time-bin state $\ket{t_m}$ is also followed by a 100 ns blank interval and attenuated to a level of $\mu$ photons per pulse at the input.

Time-bin or phase frames are transmitted alternatively, with probability $t_{tb}$ and $1-t_{tb}$, respectively; the frame rate is $R_f = 5$ MHz. Information is carried by the photon position $\ket{t_m}$ of the time-bin frames, while phase frames are used to assess the channel quality and the possible presence of an eavesdropper, as described in \cite{doi:10.1126/sciadv.1701491}.

The space-division multiplexing (SDM) system under test consists of a custom-designed modal multiplexer/demultiplexer (Cailabs PROTEUS-C-15-CUSTOM) based on multi-plane light conversion (MPLC) technology \cite{Bade:18}, capable of coupling 15 Hermite-Gaussian modes $HG_{np}$ into an 8 km long few-mode fiber (FMF, Prysmian GI-LP9) \cite{Sillard:16}, and decomposing them at the output. The multiplexer/demultiplexer (MDM) consists of cascaded phase plates that convert $M$ Gaussian beams from single-mode fibers into $M$ FMF modes. These reciprocal devices also decompose $M$ input FMF modes into $M$ Gaussian beams coupled to output single-mode fibers. Thus, the MDM operates passively, even at the single-photon level.
The fiber supports $M = 15$ modes per polarization, grouped into $Q = 5$ quasi-degenerate mode groups.
Degenerate modes are characterized by same modal propagation constant $\beta_j$, $j=1,..,Q$. Modal group $j$ includes $j$ modes; specifically, group modes are $HG_{00}$ for $j=1$, $HG_{01}$, $HG_{10}$ for $j=2$, $HG_{02}$, $HG_{11}$, $HG_{20}$ for $j=3$, $HG_{03}$, $HG_{12}$, $HG_{21}$, $HG_{30}$ for $j=4$, and $HG_{04}$, $HG_{13}$, $HG_{22}$, $HG_{31}$, $HG_{40}$ for $j=5$.

The linear transmission properties of the SDM system were characterized with 40 m or 8 km of FMF spliced to the MDM. Low-power, 70 fs pulses at 1550 nm were injected into each modal channel, and the output power was measured. The insertion loss $IL_{np}$ for each mode was defined as the ratio of total output power to the input power coupled into mode $(n, p)$. The average insertion loss at 1550 nm was –8.4 dB (–12.4 dB) for 40 m (8 km) of FMF, and the worst case was –10.6 dB (–14.2 dB). By summing the powers of quasi-degenerate modes, group-wise insertion losses $IL_j$ were obtained and reported in Table \ref{tab:IL_table}.
Modes within a same group and among different groups exchange power as a consequence of random mode-coupling (RMC) in the FMF \cite{Gloge:6774107,Olshansky:75,Savovi2019PowerFI,Zitelli:10848169}, caused by fiber imperfections, micro and macro bending. Degenerate modes exchange power rapidly; therefore, power equipartition can be assumed within each group. Between different groups, the RMC produces two asymmetric power flows toward lower- or higher-order groups, favoring the lower-order ones. The modal cross-talk was measured by adding the powers of these quasi-degenerate modes at the input and at the output. The results for the group-by-group cross-talk are collected in Table \ref{tab:XT_table} after 8 km of FMF \cite{Zia:10799905}. In this section, cross-talk measurement will be repeated using single-photon pulses, therefore demonstrating the ergodicity of the RMC process \cite{Zitelli:10848169}.

The insertion loss after 40 m in Table \ref{tab:IL_table} indicate that each MDM introduces a 3.5 dB loss. FMF attenuation is comparable to the standard single-mode fibers and is typically 0.25 dB/km at 1550 nm; hence, the impact of MDM becomes negligible in QKD systems that employ several tens of kilometers of FMF.

At the input of the modal multiplexer, an optical splitter with delay lines produces the quantum signals coupled to the FMF modes. The signal power, measured by the single-photon detector $D_{IN}$ (ID Qube NIR gated), was set to $\mu \simeq 2.5$ photons/frame, corresponding to $\simeq 1$ photon/frame at the FMF input. Based on Table \ref{tab:IL_table}, the mean photon number at the receiver is approximately $\mu_o = 0.15$ photons/frame.

\begin{table}[h]
    \centering
    \begin{tabular}{cccccc}
        \hline
        Distance & Group 1 & Group 2 & Group 3 & Group 4 & Group 5 \\
        \hline
        40 m & -7.14 & -7.21 & -7.84 & -8.72 & -9.18 \\
        \hline
        8 km & -12.66 & -12.42 & -12.46 & -12.06 & -12.67 \\
        \hline
    \end{tabular}
    \caption{Insertion Loss (dB) after 40 m or 8 km of FMF}
    \label{tab:IL_table}
\end{table}

\begin{table}[h]
    \centering
    \begin{tabular}{cccccc}
        \hline
        Input: & Group 1 & Group 2 & Group 3 & Group 4 & Group 5 \\
        \hline
        Out group 1 & -0.67 & -14.23 & -19.49 & -21.61 & -23.47 \\
        \hline
        Out group 2 & -10.82 & -1.03 & -11.71 & -15.81 & -18.31 \\
         \hline
        Out group 3 & -15.25 & -10.15 & -1.46 & -10.56 & -14.45 \\
         \hline
        Out group 4 & -17.45 & -13.45 & -8.50 & -1.66 & -8.51 \\
         \hline
        Out group 5 & -18.90 & -14.89 & -11.87 & -7.06 & -0.95 \\
        \hline
    \end{tabular}
    \caption{Cross-talk (dB) after 8 km of FMF}
    \label{tab:XT_table}
\end{table}

In the transmission experiment, three signals comprising repeated $\ket{\phi}$ or $\ket{t_m}$ states were sent: signal $A$ into mode $HG_{00}$ (group 1), signal $B$ into mode $HG_{11}$ (group 3), and signal $C$ into mode $HG_{22}$ (group 5). Signal $B$ was delayed by 100 ns via 20 m of single-mode fiber before entering the multiplexer. Signals $A$, $B$, and $C$ were transmitted through 8 km of FMF, experiencing insertion loss and intermodal crosstalk (XT). 

At the receiver, Bob collects individual modes $HG_{np}$. An optical switch $S$ directs signals to the time-bin detector $D_T$ or the phase detector $D_P$. Detector $D_T$ (ID Qube NIR gated) measures the arrival time $t_m$ with respect to the FPGA trigger gate $G$, using a histogram counter (IDQ ID1000) with 25 ps resolution. Detector $D_P$ measures the output of a free-space time-delay interferometer formed by two beam splitters (BS) and mirrors (M). The interferometer overlaps two pulse replicas delayed by $T_p$; a piezo-mounted mirror (PZ) adjusts the phase $\phi_B$ to achieve destructive interference. A polarization controller (Pol. Cntr.) ensures horizontal polarization at the interferometer input. A secondary detector $D'_P$ may be used for phase state discrimination: phase bit 0 yields a click on $D'_P$ and none on $D_P$; bit 1 results in a click on $D_P$ only. Thus, the interference contrast is assessed by comparing $D_P$ counts with and without interference.

Detector $D_{IN}$ continuously monitors the input photon flux. The input counts for signals $A$, $B$, and $C$ are balanced within a 5 \% margin.

The detectors, with efficiency $\eta = 0.15$ and dead time $t_d = 100$ ns, are not limited by dead time in this regime. 
Instead, sensors dead time is nested into a blank time interval 
used to suppress crosstalk from $A$ and $C$ to $B$ through time-windowing. Signals $A$, $B$, $C$ are coupled to non-adjacent modal groups in order to reduce the cross-talk; although this is sufficient to limit cross-talk between signals $A$ and $C$ after the modal demultiplexer, it does not appear sufficient to isolate signal $B$ from $A$ and $C$. Hence, signals $A$ and $C$ include a 100 ns signal interval $\Delta t_1$ and a 100 ns blank interval $\Delta t_2$. Signal $B$ also includes a 100 ns blank interval, but is delayed to align with $\Delta t_2$, while $A$ and $C$ occupy $\Delta t_1$. Detectors for modal groups 1 or 5 are gated on $\Delta t_1$ and experience dead-time on $\Delta t_2$; the opposite if group 3 must be detected. This allows time-gated detection of $B$ during $\Delta t_2$, effectively isolating it from crosstalk by $A$ and $C$.

\begin{figure*}
\centering
\includegraphics[width=0.6\textwidth]{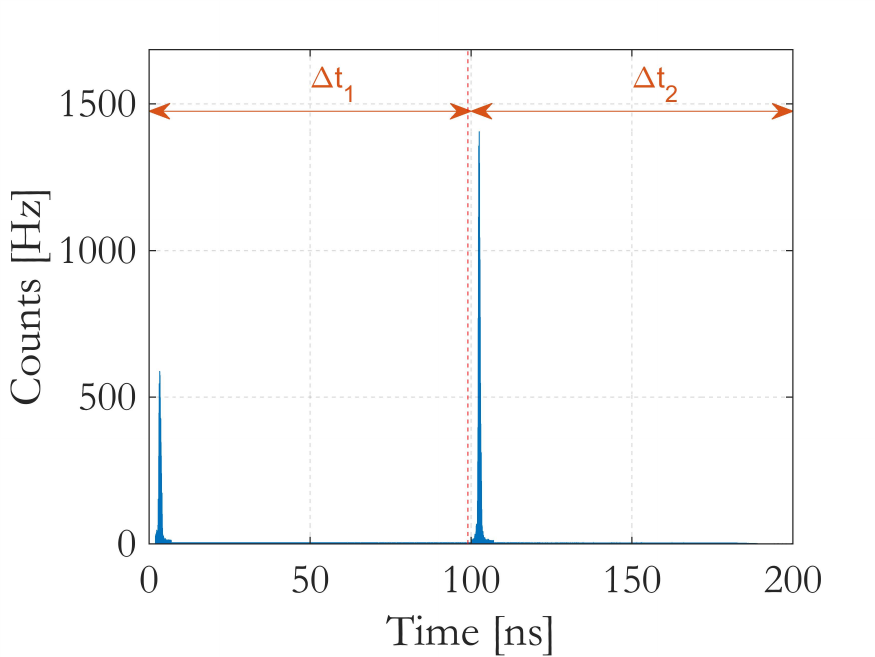}
\caption{Histogram of time-bin state received from mode $HG_{11}$; here the received pulses occupy time slot 1 into the frame. The received signal $B$ occupies the time interval $\Delta t_2$ while signals $A$ and $C$ appear in $\Delta t_1$.}
\label{fig:In9_8_4_Delay8_1pulse}
\end{figure*}

\begin{figure*}
\centering
\includegraphics[width=0.6\textwidth]{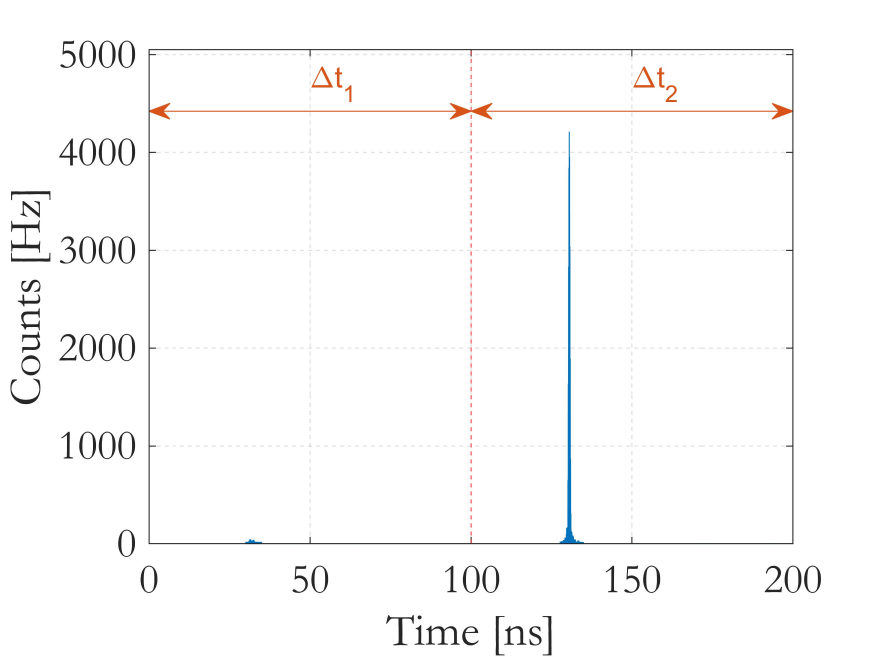}
\caption{Histogram of time-bin state received from mode $HG_{00}$; the received pulses occupy time slot 20 into the frame. The received signal $A$ occupies the time interval $\Delta t_2$ while crosstalk from $C$ in $\Delta t_1$; signal $B$ is disconnected.}
\label{fig:In9_4_Delay9_1pulse}
\end{figure*}

Figure \ref{fig:In9_8_4_Delay8_1pulse} shows a histogram from $D_T$ connected to output mode $HG_{11}$; here the received pulses occupy time slot 1 into the frame. The received signal $B$ occupies the time interval $\Delta t_2$ while crosstalk from $A$ and $C$ appears in $\Delta t_1$ but is removed using time windowing. To quantify crosstalk between $A$ and $C$, $B$ was disconnected and $A$ was delayed by 100 ns. Figure \ref{fig:In9_4_Delay9_1pulse} shows negligible XT from $C$ in $\Delta t_1$ relative to delayed $A$ in $\Delta t_2$. This confirms that signals $A$ and $C$ can share a time window without significant crosstalk, enabling only $B$ to be delayed.

\begin{figure*}
\centering
\includegraphics[width=0.6\textwidth]{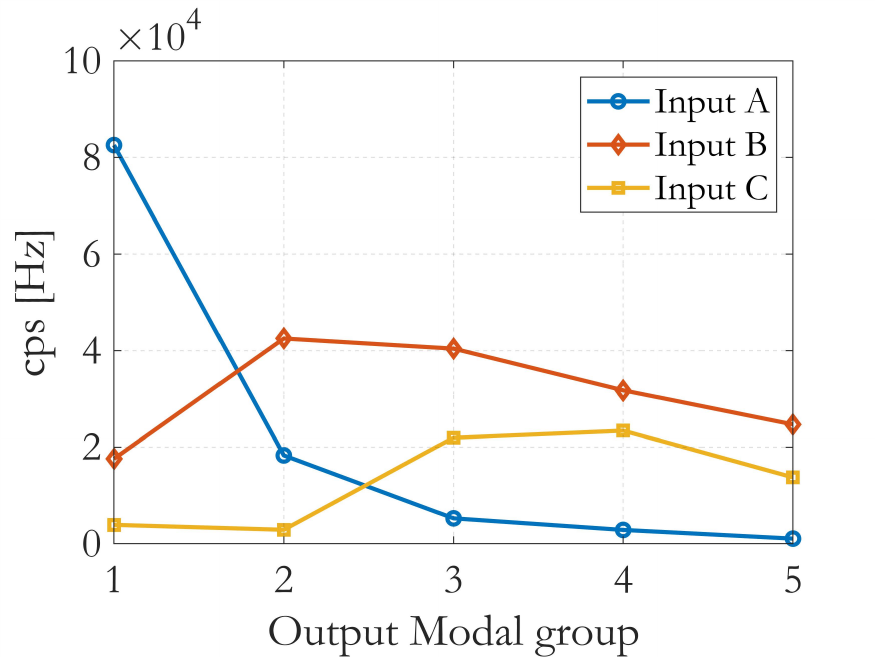}
\caption{Modal output counts per second summed group-wise, originating from signals $A$, $B$ and $C$.}
\label{fig:In9_8_4_Signal}
\end{figure*}

\begin{figure*}
\centering
\includegraphics[width=0.6\textwidth]{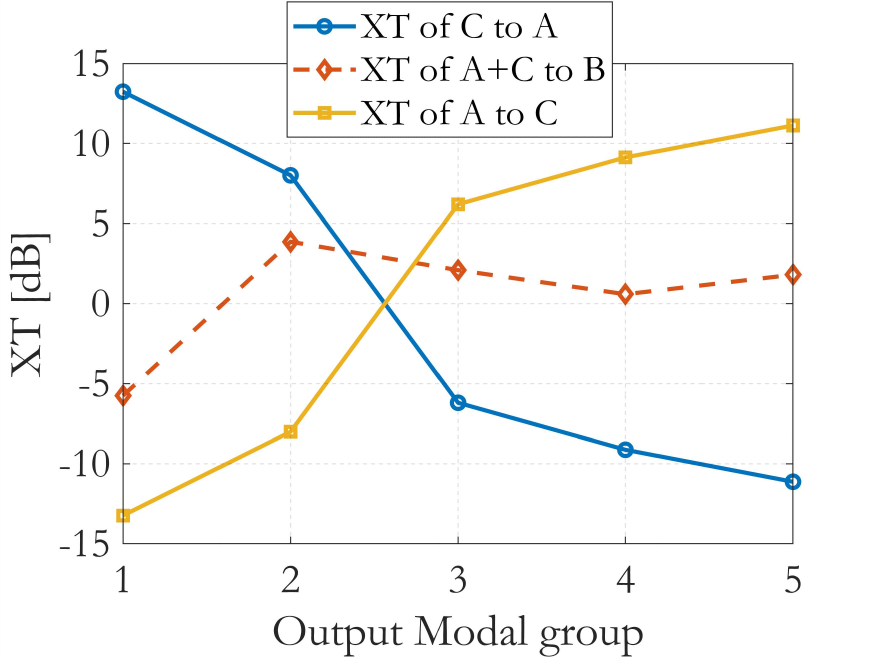}
\caption{Modal output cross-talk summed group-wise, originating from signals $A+C$ to $B$ or between $A$ and $C$. XT of $A+C$ to $B$ is represented as a dotted line because it is completely eliminated by time windowing. By connecting the detector to the individual output modes, $A+C$ ($B$) was measured from the time bit slot of the pulse in $\Delta t_1$ ($\Delta t_2)$).}
\label{fig:In9_8_4_XT}
\end{figure*}

Signal $B$'s count rate and XT were evaluated by applying $A$, $B$, and $C$, delaying $B$, and measuring $B$ in $\Delta t_2$. For signals $A$ and $C$, $B$ was disconnected, and $A$ was delayed. Detector was connected to the individual modal outputs, and outputs were summed group-wise, a technique known as modal group demultiplexing. Figure \ref{fig:In9_8_4_Signal} shows the resulting counts per second (cps) originating from for time-bin signals $A$, $B$ and $C$, for different output modal groups. Figure \ref{fig:In9_8_4_XT} is the XT from $A+C$ to $B$ and between $A$ and $C$, calculated as the count ratio in $\Delta t_2$ and $\Delta t_1$ for time-bin signals ($XT=10\log_{10}[c_i(\Delta t_2)/c_i(\Delta t_1)]$).
For example, cross-talk of $A+C$ to $B$ was measured from the single-pulse histograms (like Fig. \ref{fig:In9_8_4_Delay8_1pulse}), counting detections of $A+C$ on the time-bin slot of the pulse in $\Delta t_1$ and $B$ on the corresponding time-bin in $\Delta t_2$. For cross-talk of $A$ to $C$ it was disconnected signal $B$ and delayed $A$, so that $C$ ($A$) counts were falling on interval $\Delta t_1$ ($\Delta t_2$).
From Fig. \ref{fig:In9_8_4_Signal} we note that signal photons flow from single input modes to any output modal group, as a consequence of random mode-coupling (RMC) in the FMF \cite{Gloge:6774107,Olshansky:75,Savovi2019PowerFI,Zitelli:10848169}; signal $A$ remains confined to group 1, signal $B$ flows from input group 3 to output 2, 3, 4, and signal $C$ from input group 5 to output 3 to 5.

Noise floor counts were observed due to the finite extinction ratio of the IM. These were measured in pulse-absent frame intervals and rescaled to 100 ns. The signal-to-noise ratio (SNR) was computed as the ratio of counts in the 1.54 ns pulse slot to floor counts over 100 ns. For $A$, $B$, and $C$, the average SNR was 11.3 dB, corresponding to a false-detection probability $p_{SNR} < 0.075$.

In Fig. \ref{fig:In9_8_4_XT}, the dotted line shows that crosstalk from $A+C$ to $B$ was fully removed by time windowing. However, XT between $A$ and $C$ (co-timed) remained above 11.1 dB, implying a false-detection probability $p_{XT} < 0.08$.

To evaluate phase state performance, sequences of $d = 64$ pulses with phase difference $\phi_A = \pi$ were transmitted, and $\phi_B = 0$ was set at the receiver. Counts $c_i(\Delta t_1)$ and $c_i(\Delta t_2)$ (with interference) were compared with $c_0(\Delta t_1)$ and $c_0(\Delta t_2)$ (with one interferometer arm blocked).

\begin{figure*}
    \centering
    \includegraphics[width=0.7\textwidth]{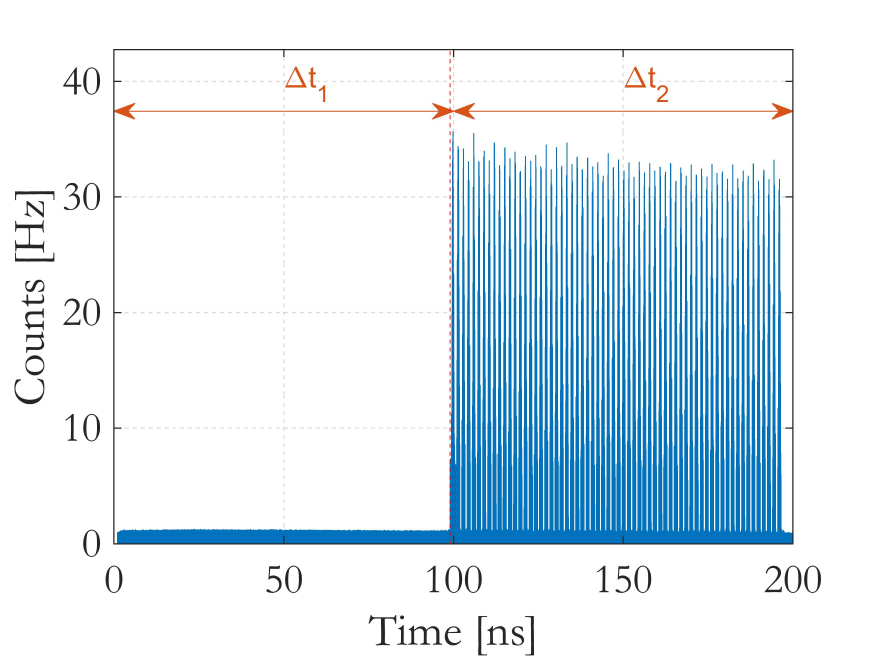}
    \caption{Received histogram for output mode $HG_{00}$, when phase states are transmitted and in the non-interfering case. Signal $A$ was delayed and $B$ disconnected.}
    \label{fig:In9_4_Delay9_64pulse_1arm}
\end{figure*}

\begin{figure*}
    \centering
    \includegraphics[width=0.7\textwidth]{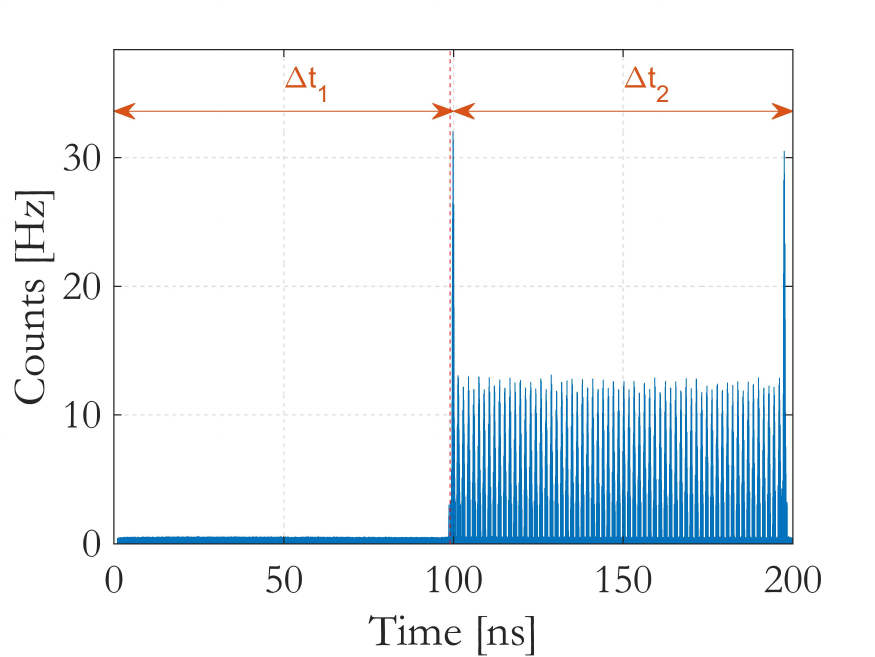}
    \caption{Received histogram for output mode $HG_{00}$, corresponding to Fig. \ref{fig:In9_4_Delay9_64pulse_1arm}, when phase states are transmitted and in the interfering case.}
    \label{fig:In9_4_Delay9_64pulse_2arm}
\end{figure*}

\begin{figure*}
    \centering
    \includegraphics[width=0.7\textwidth]{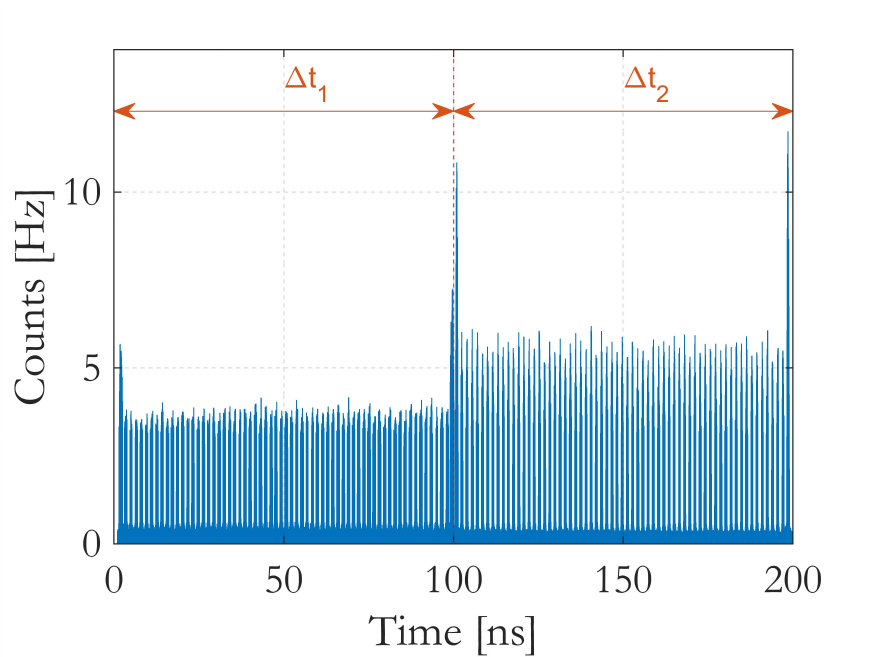}
    \caption{Received histogram for output mode $HG_{11}$, when phase states are transmitted and in the interfering case. The delayed signal $B$ is visible on time interval $\Delta t_2$ and the XT from $A+C$ on $\Delta t_1$.}
    \label{fig:In9_8_4_Delay8_64pulse_2arm}
\end{figure*}

Figure \ref{fig:In9_4_Delay9_64pulse_1arm} shows the non-interfering case for output mode $HG_{00}$, where $A$ was delayed and $B$ disconnected. Figure \ref{fig:In9_4_Delay9_64pulse_2arm} shows the interference pattern with both interferometer arms active; in the figure, the two external pulses do not interfere, while the inner ones show reduced power due to destructive interference. Figure \ref{fig:In9_8_4_Delay8_64pulse_2arm} is a second example of the interfering case for output mode $HG_{11}$ when the three signals are transmitted, with $B$ visible on $\Delta t_2$ and $A+C$ on  $\Delta t_1$.

The extinction ratio for $B$ is defined as:

\begin{equation}
ER_{\phi} = 10 \log_{10} \left( \frac{2 c_0(\Delta t_2)}{c_i(\Delta t_2)} \right) .
\label{eq:ER}
\end{equation}

\begin{figure*}
    \centering
    \includegraphics[width=0.6\textwidth]{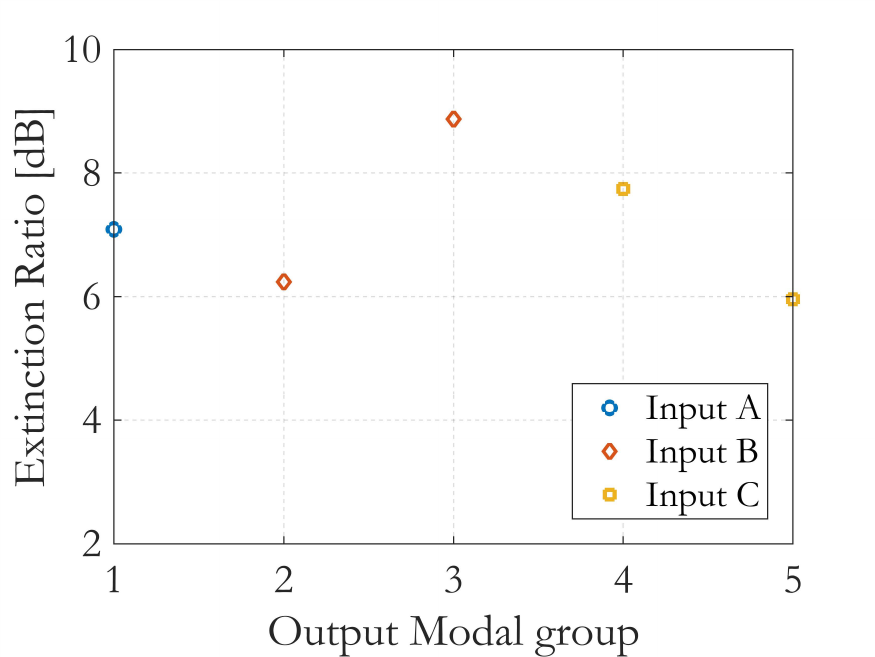}
    \caption{Extinction ratios between non-interfering and interfering phase states, measured from output signals summed group-wise, and originating from input signals $A$, $B$ and $C$.}
    \label{fig:ERbyGroups}
\end{figure*}

\begin{figure*}
    \centering
    \includegraphics[width=0.6\textwidth]{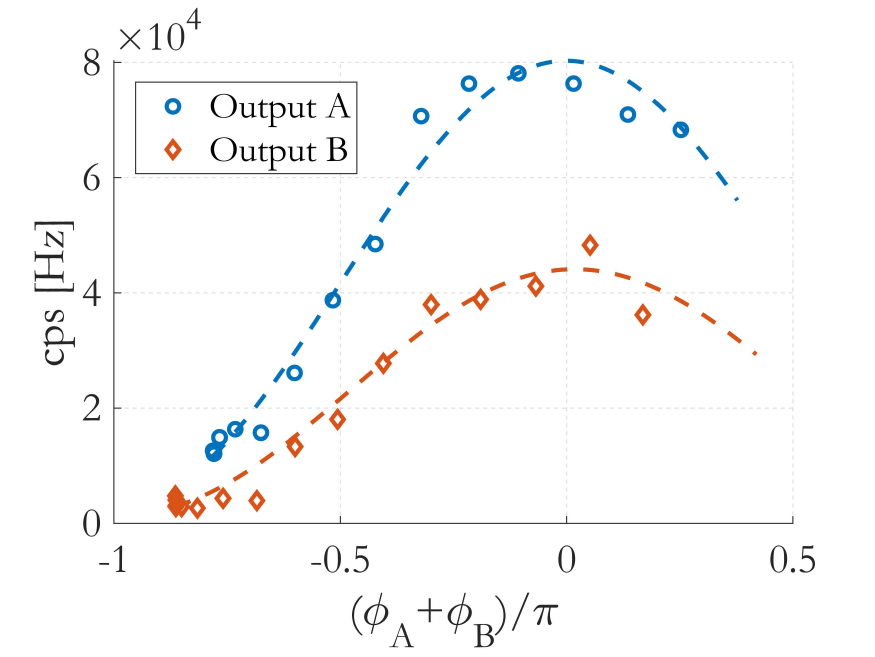}
    \caption{Counts for phase frames of signals $A$ and $B$ at the output of the unbalanced interferometer, against the total phase $\phi_A+\phi_B$ between adjacent pulses. Dotted lines are the fits from Eq. \ref{eq:CountsPhase}.}
    \label{fig:CountsVsPhase}
\end{figure*}

In a similar way it is measured the $ER$ on $A$ when this is delayed on $\Delta t_2$ and $B$ disconnected. In Fig. \ref{fig:ERbyGroups} (blue circle) $ER$ is shown for signal $A$. Extinction ratios were measured from output signals summed group-wise, considering only the contribution of $A$ to group 1, $B$ to groups 2,3 and $C$ to groups 4,5; it ranged from 5.9 to 8.9 dB (Fig. \ref{fig:ERbyGroups}), indicating a mean detection suppression factor of 5.4 due to interference, or that the probability of detection of wrong phase states was $p_\phi = 0.18$ assuming unitary probability of detection without interference. 
Figure \ref{fig:CountsVsPhase} provides the measured output counts from the unbalanced interferometer when transmitting phase frames $\ket{\phi}$ with $d=64$ pulses for different values of the total phase $\phi_A+\phi_B$ among adjacent pulses; here, $\phi_A$ is the differential phase from the PM at the transmitter and $\phi_B$ the phase added by the unbalanced interferometer. Counts are fitted by the interference sinusoid.

\begin{equation}
I(\phi) = I_0 \big[1+V_{ij}\cos(\phi_A+\phi_B)\big]    ,
\label{eq:CountsPhase}
\end{equation}

with $V_{ij}$ the visibility between adjacent pulses $i$ and $j$. Details are discussed in the next section.

\section{\label{sec:Discussion}Discussion and Conclusions}

In the proposed SDM system, exploiting the detector dead-time $t_d$ to suppress intermodal crosstalk (XT) requires a frame rate $R_f=1/(2t_d)$, and the pulse period is chosen to yield the desired number of qubits per pulse, $T_b=1/(2dR_f)$.
The achievable system capacity, in qubits per second, is then

\begin{equation}
C_p = Q \mu \eta R_f IL \log_2 d      ,
\label{eq:Capacity}
\end{equation}

where $IL$ is the system insertion loss (linear scale), $\eta$ the detector efficiency, $\mu$ the number of photons transmitted per frame and $Q$ the effective number of multiplexed modal groups.

Assuming an input power of $\mu'=1$ photon/frame at the input of the FMF, for both the transmitted time-bin and phase states, a frame rate of $R_f=5$ MHz, average insertion loss $IL=-12.5+4.2$ dB (excluding the input MDM), and detector efficiency $\eta=0.15$, the three signals could reach a theoretical $cps=\mu' \eta R_f IL=110.9$ kHz.
In Fig. \ref{fig:In9_8_4_Signal}, signals $A$, $B$ and $C$ detected from modal groups 1, 3, and 5 showed lower counts than this maximum.  After checking the XT values in the different output groups, better performance was obtained by reassigning group collection, collecting signal $A$ from group 1, $B$ from groups 2+3 (modes 2–6) and $C$ from groups 4+5 (modes 7–15); the measured rates for $A$, $B$ and $C$ became 82.5, 82.8, and 37.2 kcps, respectively, closely matching theory. Summing these, the SDM system yielded 202.6 kcps. Since each frame encodes 6 qubits, this corresponds to a transmission capacity $C_p=1.22$ Mqubit/s over 8 km of FMF. In practice, only a fraction $p_{tb}$ of frames carry time-bin data; the remainder $1-p_{tb}$ of frames carry phase states to detect the presence of an eavesdropper. For higher detector efficiency, the transmission capacity scales proportionally (theoretically up to 8.1 Mqubit/s for $\eta \simeq 1$), though noise will increase. Higher capacity could be reached using modal multiplexers with a higher number of modes and more advanced time-windowing techniques.

The probability of a symbol error can be assumed as the quadratic sum of the measured errors arising from the XT ($p_{XT}<0.08$) and from the floor noise ($p_{SNR}<0.075$), which equals $p_s=0.11$. XT stems from fiber RMC, and is unavoidable, whereas floor noise can be reduced using modulators with ultra-high extinction ratios. The quantum bit error rate for a pulse-position modulation format is $QBER=p_s/\log_2(d)=0.018$. 

From the above considerations and from Fig. \ref{fig:CountsVsPhase}, which shows the outputs of signals $A$ (group 1) and $B$ (groups 2+3), it is possible to extract tomography information for the time and the phase states. Time-bin states $\ket{t_j}$, are characterized by density matrix elements $\rho_{jj}=P_j$, with $j=1,..,d$ and $P_j$ the detection probability of the pulse at time $t_j$. Probabilities are obtained from the single-pulse output histograms, like Figs. \ref{fig:In9_8_4_Delay8_1pulse} and \ref{fig:In9_4_Delay9_1pulse}, by counting the detections $c_j$ corresponding to the time bin and the background $cf_j$ (floor) originating from all other time slots, hence $\rho_{jj}=c_j/(c_j+cf_j)$ for the time bin slot and $\rho_{kk}=(1-\rho_{jj})/(d-1)$ for 
all other time slots. For outputs $A$, $B$ and $C$ it was obtained $\rho_{jj}=0.93, 0.93, 0.96$ and $\rho_{kk}=0.001, 0.001, 0.0005$, respectively.

The off-axis elements of the density matrix are obtained from the curves of Fig. \ref{fig:CountsVsPhase} fitted by Eq. \ref{eq:CountsPhase}. Fits provide a visibility $V=0.93$ for both outputs $A$ and $B$. The matrix elements are then $\rho_{ij}=V_{ij}\sqrt{P_iP_j}$, with $P_i \simeq 1/d$.

In the original proposal of time-bin and phase encoding, phase states $\ket{\phi_n}$ were generated as pulse trains with differential phase $\phi_n=2\pi n/d$ \cite{doi:10.1126/sciadv.1701491}; receiver was implemented by a tree of cascaded time-delay interferometers and detectors or, in the simplest case of $n=0$, by interfering with a locally generated phase state $\ket{\phi'_0}$, which must match the Alice's $\ket{\phi_0}$ in the spatial, spectral, polarization and temporal domains. Alternatively, differential phase can implement a BB84 protocol: no local generation of phase states is needed, and the receiver is composed, as in Fig. \ref{fig:Setup}, by a single time-delay interferometer and two detectors $D_P$ and $D'_P$. Alice prepares four qubits by modulating the differential phase of the pulses with a base $X=\phi_{A1}=\{0,\pi\}$ or $Z=\phi_{A2}=\{\pi/2,3\pi/2\}$ at random, acting on the phase modulator, according to a bit sequence $a=\{a_i\}$ and a base sequence $b=\{b_i \}$; Bob adds a phase delay $\phi_{B}=\{0,\pi/2\}$ to interfering pulses, according to a base sequence $b'=\{b'_i \}$, acting on the phase control in the interferometer. The two detectors at the output of the time-delay interferometer will click alternatively and systematically if $\phi_{B}=0$ and $\phi_{A}=\{0,\pi\}$ or $\phi_{B}=\pi/2$ and $\phi_{A}=\{\pi/2,3\pi/2\}$. For other phase combinations, the two detectors could produce simultaneous clicks or no clicks, giving rise to a null bit $a'_i=\varnothing$. After raw key transmission, the bases $b, b'$ and detection instants are announced on the public channel, allowing Alice and Bob to sift the transmitted qubits and build common keys. 
An eavesdropper (Eve) can only guess the phase information adding a phase modulation $\phi_E$ at random, and re-transmits the information to Bob with a phase which does not match the Alice's information with 50\% probability. 

Using BB84 on phase frames provides robustness against limited extinction ratios between different interference patterns. In the tested system, it was measured an average extinction ratio $ER=7.3$ dB between destructive interference and no interference, corresponding to a probability of error in detecting phase states $p_{\phi}=0.18$. Sifting techniques and privacy amplification, typical of BB84 and similar protocols, can correct such errors for sufficiently long keys. 
Specifically, using Eqs. 2 and 3 in \cite{Tomamichel2012}, with security and correction rates $\epsilon_{sec}=\epsilon_{corr}=1\times10^{-14}$, channel error tolerance $Q_{tol}=0.01$, equal number of raw bits $n$ and bits for parameter estimation $k$, expected number of exchanged qubits $M=n$ (until $n$ raw bits and $k$ bits for parameter estimation are gathered), preparation quality $q=1$ and robustness (probability that protocol aborts in the absence of an eavesdropper) $\epsilon_{rob}=p_{\phi}=0.18$, the expected key rate normalized to $n$ is $r \geq 0.64$ for $n \geq 10^6$.
In this scheme, phase frames are used as qubits for eavesdropper detection, while time-bin frames carry key data; phase frames are sent with the minimum fraction $1-p_{tb}$ required for security estimation.
The choice of $p_{tb}$ is limited to the minimum value necessary to assess in real time the generation of correct sifted key, which are used here to certify the absence of an eavesdropper.

\section{Acknowledgments/Funding}

The author wishes to thank Fabio Sciarrino, Gonzalo Carvacho and Danilo Zia for fruitful discussions.

Project ECS 0000024 Rome Technopole, Funded by the European Union – NextGenerationEU



\end{document}